**Social Statements: A Proposal for a Social-Value Balance Sheet and Profit–Loss Statement**

Takeshi Kato[1*], Yoshinori Hiroi[2], Sae Horiguchi[3], Tetsushi Koike[3], Shingo Hashimoto[3]

[1] Strategy Development Office, Strategy and Administration Department, Institutional Advancement and Communications, Kyoto University
[2] Professor Emeritus of Kyoto University
[3] Dynax Urban Environment Research Inst., Inc.

[*] Corresponding author
Email: kato.takeshi.3u@kyoto-u.ac.jp
ORCiD: https://orcid.org/0000-0002-6744-8606

## Abstract

This study proposes a new set of a firm's "social statements" that represent social value, in contrast to conventional financial statements that represent economic value. Financial statements externalize social and environmental costs, and this externalization is one of the primary causes of contemporary social problems. Insights from anthropology, philosophy, and sociology suggest that social value is grounded in social relationships, joint actions, and communication. Building on this understanding, we assign numerical indicators of a firm's social relationships with external stakeholders to the items of a balance sheet and a profit–loss statement as social statements. This approach enables unified measurement units and simplified calculation compared with existing methods for evaluating social impact or social value. Moreover, similar to financial statements, social statements allow firms to be assessed using managerial indicators such as equity ratios and profit margins. The significance of social statements lies in incorporating social value—alongside financial value—into corporate decision-making, and in encouraging social transformation as firms publicly articulate their social value.

# 1. Introduction

Wealth inequality and resource waste have become major social issues worldwide. Conversely, enhancing well-being and sustainability constitutes a grand challenge. According to the World Inequality Report 2022, the top 10% of wealth holders possess 76% of global wealth, while the top 1% alone holds 38% [1]. Resource consumption continues to increase alongside global GDP growth, far exceeding the planet's sustainable material footprint [2]. In an extreme interpretation, these social problems stem from the enslavement of labor and the colonization of resources within capitalist enterprises [2].

One of the primary factors contributing to social problems is the set of financial statements—namely, the balance sheet and the profit–loss statement—that serve as representative managerial indicators for corporations [3, 4]. Specifically, the balance sheet (B/S) does not record intangible assets such as brand value, customer bases, or organizational capabilities, and human capital is treated as a cost. In the profit–loss statement (P/L), social and environmental costs—including labor exploitation, burdens on local communities, and environmental pollution—are externalized, while investments in research and development or sustainability for the future are recorded as expenses. Consequently, although financial statements function as effective indicators for pursuing short-term corporate profits, they fail to represent social value or environmental value.

To compensate for these shortcomings of financial statements, the disclosure of non-financial information by corporations has been expanding [5, 6]. In Europe, the Corporate Sustainability Reporting Directive came into effect in 2023, and in Japan, disclosures related to sustainability and human capital have been required in securities reports since fiscal year 2024. Specifically, as ESG management indicators, environmental indicators such as greenhouse gas emissions and resource use, social indicators such as gender wage gaps and working hours, and governance indicators such as the ratio of outside directors and ethical conduct guidelines have been introduced. Among these indicators, while environmental indicators assess the impact on the external environment of the firm, social indicators remain limited to evaluating internal working conditions rather than the external society. Governance indicators, moreover, inherently concern internal corporate control and do not involve external stakeholders.

What, then, is the social value that social indicators ought to represent? According to anthropologist Graeber's *Theory of Value* [7], "*value becomes the way people represent the importance of their own actions to themselves: normally, as reflected in one or another socially recognized form,*" and "*value is ultimately the power to create social relations*." Philosopher Deguchi argues that value is what our embodied actions aim toward, a vector that gives clear direction to our behavior, and an adhesive that enables joint action [8, 9]. Furthermore, in sociologist Luhmann's theory of social systems [10], a social system is composed of a network whose elements are communications; thus, social value can be understood as grounded in communication. Taken together, the views of Graeber, Deguchi, and Luhmann suggest that social indicators representing a firm's social value should reflect its social relationships, joint actions (such as production, distribution, and consumption), and communications (interactions) with external stakeholders.

The Cooperative Business Survey Report [11] introduces several firms that conduct cooperative business while adopting management indicators that emphasize social and environmental value. For example, Ame Kaze Taiyo, Inc. has proposed "social financial statements" [12], in which the balance sheet (B/S) assigns the amount of corporate vision communicated (in person-hours) to social capital and the amount of external stakeholders' empathy toward the vision (in number of persons) to social assets, while the profit–loss statement (P/L) incorporates impact indicators such as the monetary value of transparent transactions between producers and consumers, the number of communications, and the number of days spent in regions different from one's place of residence. Saka no Tochu, Inc. uses the number of continuing customers and the number of partner producers as impact indicators [13]. Fermenstation Co., Ltd. has proposed an impact model aligned with the development of its business phases, identifying emitters, users, and consumers of unused resources as sources of impact [14]. These examples share a common aim: to measure the social relationships and joint actions between firms and external stakeholders, as discussed earlier. However, challenges remain. In Ame Kaze Taiyo's social financial statements, the measurement units differ between the B/S and P/L, making their relationship less clear compared with conventional financial statements. Moreover, although each firm's impact indicators reflect its own social value, they are less likely to become generalized social indicators comparable to existing financial statements.

Building on the above, this study aims to propose a new set of social statements that can serve as a counterpart to economic financial statements. To represent a firm's social value, we construct social statements from indicators that reflect the social relationships—joint actions and communications—between the firm and external stakeholders, and present them in the same format as financial statements, namely a balance sheet (B/S) and a profit–loss statement (P/L). Specifically, in the Methods section of this report, we explain each item of the social statements in comparison with conventional financial statements and introduce managerial indicators (such as equity ratios and return on assets) derived from the social statements. In the Calculation Examples section, we represent the social relationships between the firm and external stakeholders using a network graph and present the resulting social statements and managerial indicators computed from it. In the Discussion section, we outline the advantages and remaining issues of the proposed social statements and describe potential future developments.

# 2. Methods

## 2.1 Social Statements

In the social statements, the balance sheet (B/S) is considered to cover the full period, while the profit–loss statement (P/L) pertains to the current period. We assign to each B/S and P/L item the social relationships—joint actions and communications—between the firm and external stakeholders. Among these social relationships, we distinguish those between the firm and residents (such as producers, consumers, and customers), between partners and residents, and between local governments and residents. The numerical values of social relationships are not treated as cumulative counts or durations of actions and communications; rather, they are modeled using saturation functions of frequency or time. For example, in cases where long-term trust relationships exist, a saturation value is assigned. This is because it is intuitively implausible for the numerical representation of trust to increase indefinitely. Tables 1 and 2 present the correspondence between financial statements and social statements for the B/S and P/L, respectively.

Table 1. Correspondence Between Financial Statements and Social Statements (Balance Sheet).

| Item | | Financial Statements | Social Statements |
|---|---|---|---|
| Assets | Current Assets | Assets convertible into cash within one year | Unsaturated relationships |
| | Fixed Assets | Assets requiring more than one year to convert into cash | Saturated relationships |
| Liabilities | Current Liabilities | Liabilities to be repaid within one year | Unsaturated relationships between partners or local governments and residents |
| | Fixed Liabilities | Liabilities to be repaid over more than one year | Unsaturated relationships between partners or local governments and residents |
| Net Assets | Equity Capital | Capital that does not require repayment | Relationships between the firm itself and residents |
| | Retained Earnings | Profit for the current period | Relationships newly developed during the period between the firm and partners, local governments, or residents |

Table 2. Correspondence Between Financial Statements and Social Statements (Profit–Loss Statement).

| Item | | Financial Statements | Social Statements |
|---|---|---|---|
| Revenue | Cost of Goods Sold | Costs required to generate revenue | Building relationships with residents through partners |
| | Selling, Gen. and Admin. Expenses | Expenses necessary for sales and operations | Building relationships between the firm and partners, local governments, or residents through public communication |
| | Non-Operating Expenses | Costs associated with activities outside the core business | Maintaining relationships between the firm and partners or local governments |
| | Extraordinary Losses | Costs arising from exceptional events | Loss of relationships due to disasters, corporate relocation, and similar events |
| | Corporate Taxes | Taxes paid to local governments | Building relationships with residents mediated by local governments |
| | Net Income | Final profit after deducting all costs | Developing relationships between the firm and residents during the current period |

Table 3 presents the correspondence between the balance sheet and profit–loss statement within the social statements. The main differences from financial statements are as follows. First, in the social statements, cost of goods sold and corporate taxes correspond to partners and local governments, respectively, making these two items closely related. Second, whereas retained earnings and net income coincide in financial statements, in the social statements retained earnings include selling, general and administrative expenses and non-operating expenses in addition to net income. Details are provided in the Calculation Examples section.

Table 3. Correspondence Between the Balance Sheet and Profit–Loss Statement in the Social Statements.

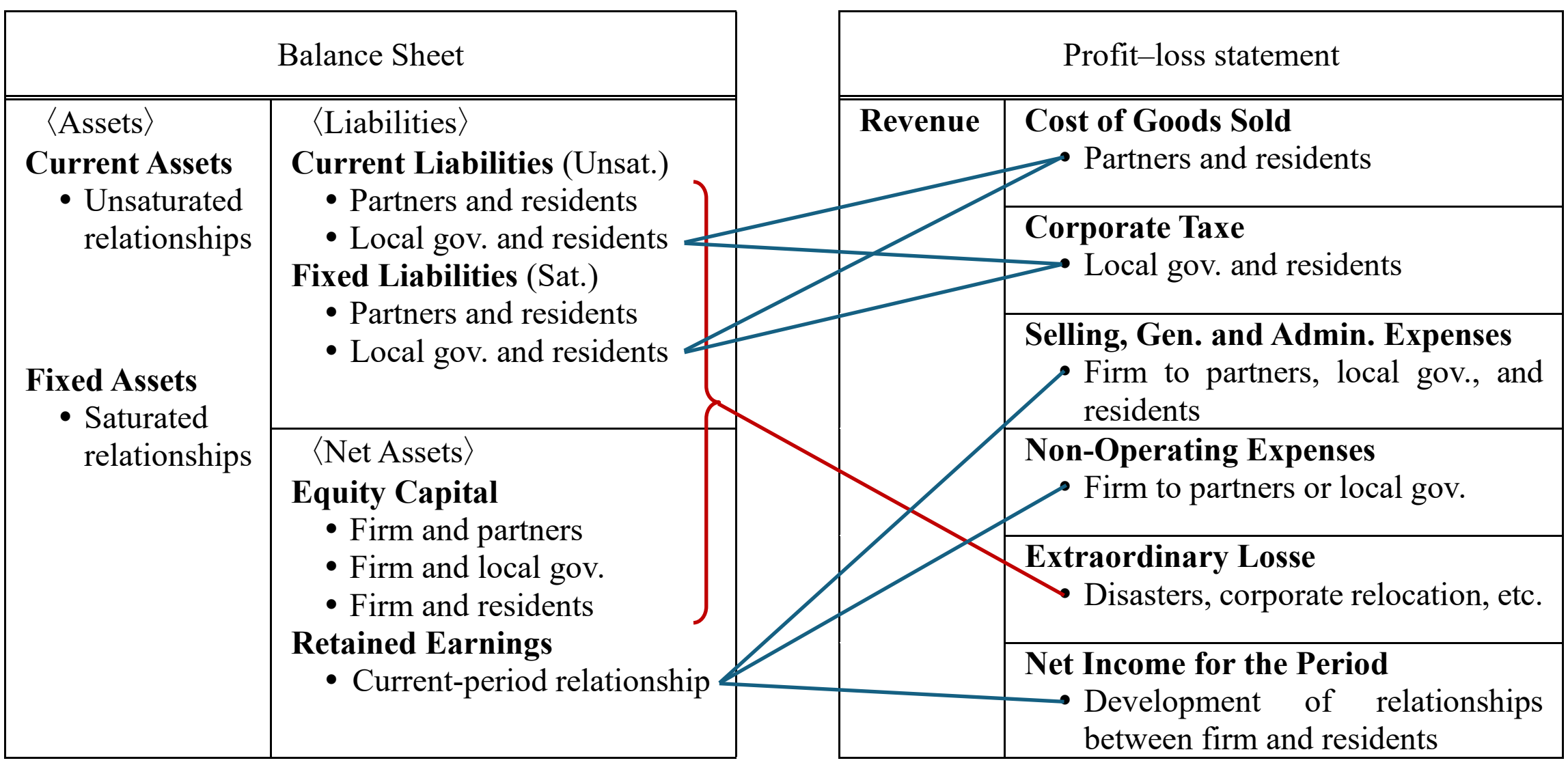

## 2.2 Managerial Indicators

As with financial statements, useful managerial indicators can be derived from the social statements. The formulas and meanings of representative managerial indicators are presented below.

- **Balance Sheet**
  - **Equity Ratio**
    - Formula: Equity Ratio = Net Assets / Total Assets
    - Meaning: Indicates low dependence on others (partners or local governments) and high stability.

- **Current Ratio**
  - Formula: Current Ratio = Current Assets / Current Liabilities
  - Meaning: Suggests that the firm is generating social returns through fluid, short-term activities.
- **Fixed Assets Ratio**
  - Formula: Fixed Assets Ratio = Fixed Assets / Net Assets
  - Meaning: Indicates the establishment of stable social relationships (trust relationships).
- **Debt Ratio**
  - Formula: Debt Ratio = Liabilities / Net Assets
  - Meaning: Indicates a high degree of dependence on others.

■ **Profit–Loss Statement**

- **Net Profit Margin**
  - Formula: Net Profit Margin = Current Profit / Revenue
    where
    Current Profit = Revenue − Cost of Goods Sold − Corporate Taxes − Extraordinary Losses
    = Net Income + Selling, Gen. and Admin. Expenses + Non-Operating Expenses
    *Note: In the social statements, the treatment of SG&A expenses and non-operating expenses differs from that in financial statements; therefore, current profit (retained earnings) is used instead of net income.*
  - Meaning: Represents the firm's overall capacity to generate social returns.
- **Return on Assets (ROA)**
  - Formula: ROA = Current Profit / Total Assets
  - Meaning: Indicates the extent to which total assets are effectively utilized to generate social returns.
- **Return on Equity (ROE)**
  - Formula: ROE = Current Profit / Net Assets (Equity Capital)
  - Meaning: Indicates how efficiently equity capital is used to generate social returns.
- **Return on Investment (ROI)**
  - Formula: ROI = Current Profit / Investment
    where
    Investment = Selling, Gen. and Admin. Expenses + Non-Operating Expenses
    *Note: In the social statements, relationship-building through public communication and the maintenance of existing relationships are regarded as investments.*
  - Meaning: Indicates a high recovery rate of social returns relative to investment.

Note: Because the treatment of SG&A expenses and non-operating expenses differs between social statements and financial statements, gross profit margin, operating profit margin, and ordinary profit margin are not used in the context of social statements.

## 3. Calculation Examples

In this section, we represent the social relationships between the firm and external stakeholders as a network graph and compute the social statements and managerial indicators. In the network graph, nodes represent residents, partners, or local governments, and edges (undirected and bidirectional) represent social relationships. The weight of each edge is assigned not as the cumulative total of actions or communications, but as the value of a saturation function. The numerical values for each item in the social statements are obtained by summing the weighted edges.

Examples of saturation functions include simple step functions and sigmoid functions; the latter being commonly used in simulations of human psychology. Letting $f$ denote a saturation function, one may compute $f(\sum(frequency\ or\ duration)/threshold)$, where the threshold is set either to correspond to one year—analogous to the period used in financial statements—or adjusted according to the firm's business characteristics.

The data for the social-relationship network graph and the numerical values of the social statements are derived according to the procedure described below.

- **Collect time-series data on social relationships.**
  - Node ID, Node ID, timestamp, duration, in-person/remote
  - The classification of each record into items of the social statements is determined by whether each Node ID corresponds to the firm itself, a partner, a local government, or a resident.
- **Edit the historical data.**
  - Node ID, Node ID, cumulative frequency, cumulative duration, period
  - Separate the data into full-period (for the B/S) and current-period (for the P/L).
  - If necessary, apply weighting to frequencies or durations depending on whether interactions were in-person or remote.
- **Convert the data into graph form.**
  - Node ID, Node ID, edge weight
  - Edge weights are computed using a saturation function based on cumulative frequency, cumulative duration, and period.
- **Compute the numerical values for the items in the social statements.**
  - Convert the graph data into an adjacency matrix.
  - Full-period data correspond to the B/S, and current-period data correspond to the P/L.
  - Aggregate the values of the adjacency-matrix elements corresponding to each B/S or P/L item.

### 3.1 Balance Sheet

Table 4 presents the adjacency matrix corresponding to the balance sheet. The adjacency matrix is a symmetric matrix with zeros on the diagonal, indicating that the network is an undirected graph without self-loops. Each element $a_{i,j} = a_{j,i}\ (> 0)$ represents the existence of a social relationship between nodes $i$ and $j$. The sum of the elements corresponding to direct relationships with the firm itself (orange cells), divided by two, corresponds to net assets (equity capital). The sum of the elements corresponding to relationships with residents mediated by others (partners or local governments) (green cells), divided by two, corresponds to liabilities (external capital). Elements representing relationships among partners, among local governments, between partners and local governments, or among residents (white cells) are not counted.

Table 4. Adjacency Matrix Corresponding to the Balance Sheet.

| | Firm | Ptr. 1 | Ptr. 2 | ⋯ | Gov. 1 | ⋯ | Res. 1 | Res. 2 | ⋯ | Res. n |
|---|---|---|---|---|---|---|---|---|---|---|
| Firm | 0 | $a_{o,p1}$ | $a_{o,p2}$ | ⋯ | $a_{o,g1}$ | ⋯ | $a_{o,c1}$ | $a_{o,c2}$ | ⋯ | $a_{o,cn}$ |
| Ptr. 1 | $a_{p1,o}$ | 0 | $n/a$ | ⋯ | $n/a$ | ⋯ | $a_{p1,c1}$ | $a_{p1,c2}$ | ⋯ | $a_{p1,cn}$ |
| Ptr. 2 | $a_{p2,o}$ | $n/a$ | 0 | ⋯ | $n/a$ | ⋯ | $a_{p2,c1}$ | $a_{p2,c2}$ | ⋯ | $a_{p2,cn}$ |
| ⋮ | ⋮ | ⋮ | ⋮ | ⋱ | ⋯ | ⋯ | ⋯ | ⋯ | ⋯ | ⋯ |
| Gov. 1 | $a_{g1,o}$ | $n/a$ | $n/a$ | ⋮ | 0 | ⋯ | $a_{g1c1}$ | $a_{g1,c2}$ | ⋯ | $a_{g1,cn}$ |
| ⋮ | ⋮ | ⋮ | ⋮ | ⋮ | ⋮ | ⋱ | ⋯ | ⋯ | ⋯ | ⋯ |
| Res. 1 | $a_{c1,o}$ | $a_{c1,p1}$ | $a_{c1,p2}$ | ⋮ | $a_{c1,g1}$ | ⋮ | 0 | $n/a$ | ⋯ | $n/a$ |
| Res. 2 | $a_{c2,o}$ | $a_{c2,p1}$ | $a_{c2,p2}$ | ⋮ | $a_{c2,g1}$ | ⋮ | $n/a$ | 0 | ⋯ | $n/a$ |
| ⋮ | ⋮ | ⋮ | ⋮ | ⋮ | ⋮ | ⋮ | ⋮ | ⋮ | ⋱ | ⋯ |
| Res. n | $a_{cn,o}$ | $a_{cn,p1}$ | $a_{cn,p2}$ | ⋮ | $a_{cn,g1}$ | ⋮ | $n/a$ | $n/a$ | ⋮ | 0 |

Net Assets (Equity Capital)
Liabilities (External Capital)

Total assets correspond to one-half of the sum of all matrix elements. Among these elements, one-half of the sum of the unsaturated values corresponds to current assets, while one-half of the sum of the saturated values corresponds to fixed assets. Likewise, within the liability components, one-half of the sum of the unsaturated values corresponds to current liabilities, and one-half of the sum of the saturated values corresponds to fixed liabilities. Note that retained earnings represent the portion of net assets attributable to the current period (i.e., a subset of the net-assets components).

Table 5 presents a simple numerical example of the adjacency matrix corresponding to the balance sheet (B/S). For simplicity, the saturated edge weight is set to 1 and the unsaturated edge weight is set to 0.5.

Table 5. Example of a B/S Adjacency Matrix.

| | o | p1 | p2 | g1 | c1 | c2 | c3 | c4 | c5 | c6 | c7 | c8 | c9 | c10 |
|---|---|---|---|---|---|---|---|---|---|---|---|---|---|---|
| o | 0 | 1 | 1 | 1 | 1 | 0.5 | 1 | 0.5 | 0.5 | 1 | 0.5 | 1 | 1 | 0.5 |
| p1 | 1 | 0 | 0 | 0 | 1 | 0 | 0 | 0 | 0 | 0.5 | 0 | 0.5 | 0 | 0 |
| p2 | 1 | 0 | 0 | 0 | 0 | 0 | 1 | 0 | 0.5 | 0 | 0 | 0 | 0 | 0 |
| g1 | 1 | 0 | 0 | 0 | 0 | 0.5 | 0 | 0 | 0 | 0 | 0 | 1 | 0 | 0 |
| c1 | 1 | 1 | 0 | 0 | 0 | 0 | 0 | 0 | 0 | 0 | 0 | 0 | 0 | 0 |
| c2 | 0.5 | 0 | 0 | 0.5 | 0 | 0 | 0 | 0 | 0 | 0 | 0 | 0 | 0 | 0 |
| c3 | 1 | 0 | 1 | 0 | 0 | 0 | 0 | 0 | 0 | 0 | 0 | 0 | 0 | 0 |
| c4 | 0.5 | 0 | 0 | 0 | 0 | 0 | 0 | 0 | 0 | 0 | 0 | 0 | 0 | 0 |
| c5 | 0.5 | 0 | 0.5 | 0 | 0 | 0 | 0 | 0 | 0 | 0 | 0 | 0 | 0 | 0 |
| c6 | 1 | 0.5 | 0 | 0 | 0 | 0 | 0 | 0 | 0 | 0 | 0 | 0 | 0 | 0 |
| c7 | 0.5 | 0 | 0 | 0 | 0 | 0 | 0 | 0 | 0 | 0 | 0 | 0 | 0 | 0 |
| c8 | 1 | 0.5 | 0 | 1 | 0 | 0 | 0 | 0 | 0 | 0 | 0 | 0 | 0 | 0 |
| c9 | 1 | 0 | 0 | 0 | 0 | 0 | 0 | 0 | 0 | 0 | 0 | 0 | 0 | 0 |
| c10 | 0.5 | 0 | 0 | 0 | 0 | 0 | 0 | 0 | 0 | 0 | 0 | 0 | 0 | 0 |

o: firm (self), p: partner, g: local government, c: resident

Figure 1 illustrates the network graph constructed from the B/S adjacency matrix shown in Table 5. In the figure, edges with saturated weights are drawn as thick lines, while edges with unsaturated weights are drawn as thin lines. Centered on the firm's node $o$, social relationships are formed with partners $p1$ and $p2$, the local government $g1$, and residents $c1$ through $c10$. The visualization shows that, in several cases, stable trust-based relationships have been established.

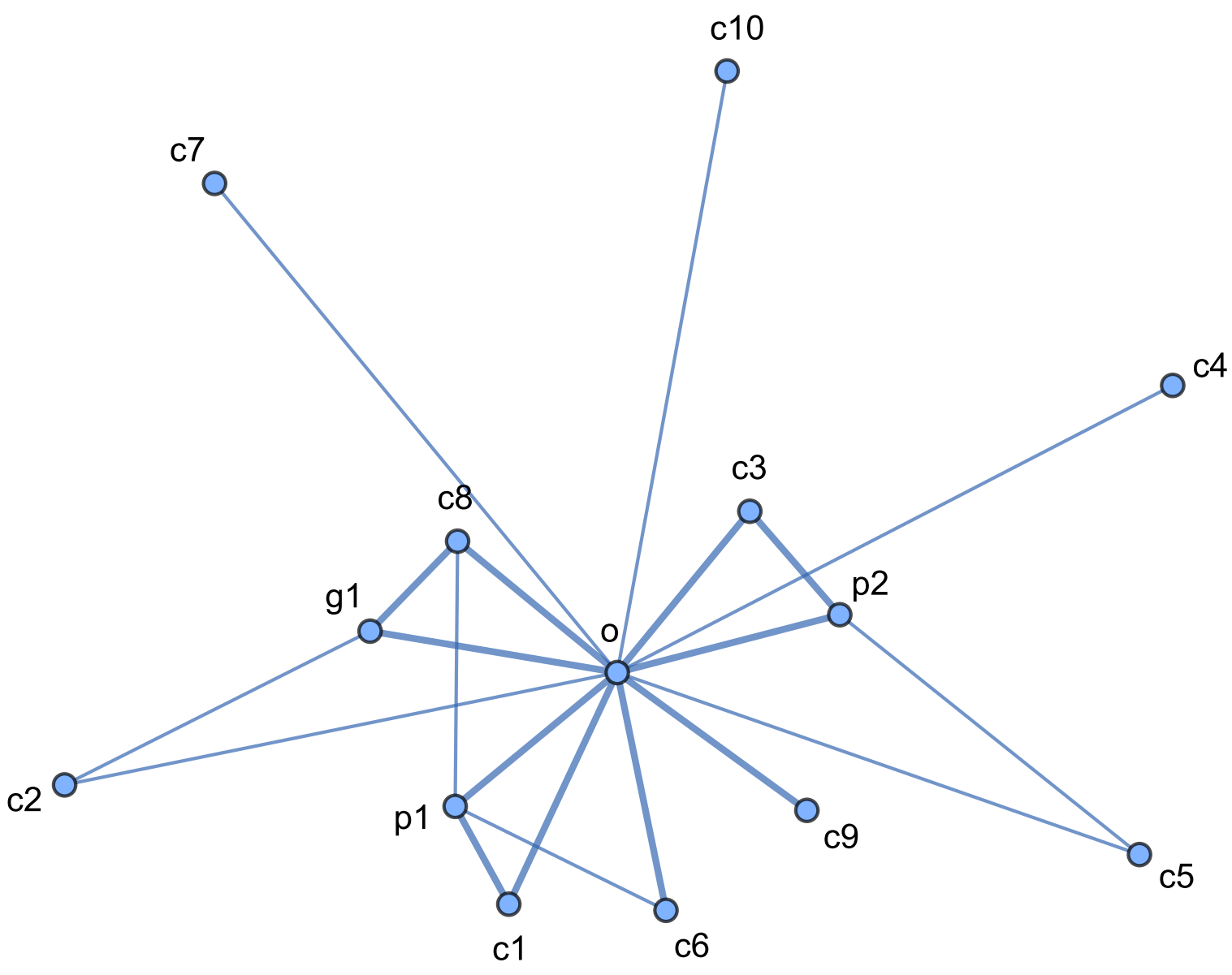

Figure 1. Network Graph Based on the B/S Adjacency Matrix.

Table 6 presents the balance sheet calculated from the B/S adjacency matrix, along with the managerial indicators derived using the formulas in Section 2.2. Although financial benchmarks cannot be directly applied, this example suggests that the firm appears to generate social value in a stable manner, given the high equity ratio and fixed assets ratio as well as the low debt ratio. Furthermore, the high current ratio indicates that the firm is also generating social returns through fluid (unsaturated) relationships.

Table 6. Balance Sheet and Managerial Indicators Calculated from the B/S Adjacency Matrix.

| Balance Sheet | | |
|---|---|---|
| Total Assets<br>15.5 | Current Assets<br>4.5 | Current Liabilities<br>2.0 |
| | Fixed Assets<br>11.0 | Fixed Liabilities<br>3.0 |
| | | Net Assets<br>(Equity Capital)<br>10.5 |

| Managerial Indicators |
|---|
| Equity Ratio<br>0.68 |
| Current Ratio<br>2.25 |
| Fixed Assets Ratio<br>1.05 |
| Debt Ratio<br>0.48 |

## 3.2 Profit–Loss Statement

Table 3 presents the adjacency matrix corresponding to the profit–loss statement. In the current period, one-half of the sum of the elements representing direct relationships formed between the firm and residents (orange cells) corresponds to net income for the period. One-half of the sum of the elements representing relationships formed through public communication between the firm and partners, local governments, or residents (yellow cells) corresponds to selling, general and administrative expenses (SG&A). One-half of the sum of the elements representing relationships maintained between the firm and partners or local governments (pink cells) corresponds to non-operating expenses. Similarly, one-half of the sum of the elements representing relationships formed with residents through partners (green cells) corresponds to the cost of goods sold, while one-half of the sum of the elements representing relationships formed with residents through local governments (light-blue cells) corresponds to corporate taxes. As in the balance sheet, elements representing relationships among partners, among local governments, between partners and local governments, or among residents (white cells) are not counted.

Table 7. Adjacency Matrix Corresponding to the Profit–Loss Statement.

| | Firm | | Ptr. 1 | Ptr. 2 | ⋯ | Gov.1 | ⋯ | Res. 1 | Res. 2 | ⋯ | Res. n |
|---|---|---|---|---|---|---|---|---|---|---|---|
| Firm | 0 | 0 | $a_{o1,p1}$ | $a_{o1,p2}$ | ⋯ | $a_{o1,g1}$ | ⋯ | $a_{o1,c1}$ | $a_{o1,c2}$ | ⋯ | $a_{o1,cn}$ |
| | 0 | 0 | $a_{o2,p1}$ | $a_{o2,p2}$ | ⋯ | $a_{o2,g1}$ | ⋯ | $a_{o2,c1}$ | $a_{o2,c2}$ | ⋯ | $a_{o2,cn}$ |
| Ptr. 1 | $a_{p1,o1}$ | $a_{p1,o2}$ | 0 | $n/a$ | ⋯ | $n/a$ | ⋯ | $a_{p1,c1}$ | $a_{p1,c2}$ | ⋯ | $a_{p1,cn}$ |
| Ptr. 2 | $a_{p2,o1}$ | $a_{p2,o2}$ | $n/a$ | 0 | ⋯ | $n/a$ | ⋯ | $a_{p2,c1}$ | $a_{p2,c2}$ | ⋯ | $a_{p2,cn}$ |
| ⋮ | ⋮ | ⋮ | ⋮ | ⋮ | ⋱ | ⋯ | ⋯ | ⋯ | ⋯ | ⋯ | ⋯ |
| Gov. 1 | $a_{g1,o1}$ | $a_{g1,o2}$ | $n/a$ | $n/a$ | ⋮ | 0 | ⋯ | $a_{g1c1}$ | $a_{g1,c2}$ | ⋯ | $a_{g1,cn}$ |
| ⋮ | ⋮ | ⋮ | ⋮ | ⋮ | ⋮ | ⋮ | ⋱ | ⋯ | ⋯ | ⋯ | ⋯ |
| Res. 1 | $a_{c1,o1}$ | $a_{c1,o2}$ | $a_{c1,p1}$ | $a_{c1,p2}$ | ⋮ | $a_{c1,g1}$ | ⋮ | 0 | $n/a$ | ⋯ | $n/a$ |
| Res. 2 | $a_{c2,o1}$ | $a_{c2,o2}$ | $a_{c2,p1}$ | $a_{c2,p2}$ | ⋮ | $a_{c2,g1}$ | ⋮ | $n/a$ | 0 | ⋯ | $n/a$ |
| ⋮ | ⋮ | ⋮ | ⋮ | ⋮ | ⋮ | ⋮ | ⋮ | ⋮ | ⋮ | ⋱ | ⋯ |
| Res. n | $a_{cn,o1}$ | $a_{cn,o2}$ | $a_{cn,p1}$ | $a_{cn,p2}$ | ⋮ | $a_{cn,g1}$ | ⋮ | $n/a$ | $n/a$ | ⋮ | 0 |

Net Income for the Period
Selling, Gen. and Admin. Expenses (SG&A)
Non-Operating Expenses
Cost of Goods Sold
Corporate Taxes

Revenue corresponds to one-half of the sum of all matrix elements. Current profit (which corresponds to retained earnings in the balance sheet) is the sum of net income for the period, selling, general and administrative expenses, and non-operating income. All matrix elements take values of zero or positive numbers. Extraordinary losses are represented by the sum of decreases in the corresponding elements from the previous period's balance-sheet adjacency matrix.

Table 8 presents a simple numerical example of the P/L adjacency matrix. As in Table 5, the saturated edge weight is set to 1 and the unsaturated edge weight is set to 0.5 for simplicity.

Table 8. Example of a P/L Adjacency Matrix.

| | o1 | o2 | p1 | p2 | g1 | c1 | c2 | c3 | c4 | c5 | c6 | c7 | c8 | c9 | c10 |
|---|---|---|---|---|---|---|---|---|---|---|---|---|---|---|---|
| o1 | 0 | 0 | 0.5 | 0 | 0 | 0 | 0 | 0.5 | 0 | 0 | 1 | 0 | 0.5 | 0 | 0.5 |
| o2 | 0 | 0 | 0 | 1 | 0.5 | 0 | 0.5 | 0 | 0 | 0 | 0 | 0.5 | 0 | 0 | 0 |
| p1 | 0.5 | 0 | 0 | 0 | 0 | 0.5 | 0 | 0 | 0 | 0 | 0 | 0 | 0.5 | 0 | 0 |
| p2 | 0 | 1 | 0 | 0 | 0 | 0 | 0 | 1 | 0 | 0 | 0 | 0 | 0 | 0 | 0 |
| g1 | 0 | 0.5 | 0 | 0 | 0 | 0 | 0 | 0 | 0 | 0 | 0 | 0 | 0.5 | 0 | 0 |
| c1 | 0 | 0 | 0.5 | 0 | 0 | 0 | 0 | 0 | 0 | 0 | 0 | 0 | 0 | 0 | 0 |
| c2 | 0 | 0.5 | 0 | 0 | 0 | 0 | 0 | 0 | 0 | 0 | 0 | 0 | 0 | 0 | 0 |
| c3 | 0.5 | 0 | 0 | 1 | 0 | 0 | 0 | 0 | 0 | 0 | 0 | 0 | 0 | 0 | 0 |
| c4 | 0 | 0 | 0 | 0 | 0 | 0 | 0 | 0 | 0 | 0 | 0 | 0 | 0 | 0 | 0 |
| c5 | 0 | 0 | 0 | 0 | 0 | 0 | 0 | 0 | 0 | 0 | 0 | 0 | 0 | 0 | 0 |
| c6 | 1 | 0 | 0 | 0 | 0 | 0 | 0 | 0 | 0 | 0 | 0 | 0 | 0 | 0 | 0 |
| c7 | 0 | 0.5 | 0 | 0 | 0 | 0 | 0 | 0 | 0 | 0 | 0 | 0 | 0 | 0 | 0 |
| c8 | 0.5 | 0 | 0.5 | 0 | 0.5 | 0 | 0 | 0 | 0 | 0 | 0 | 0 | 0 | 0 | 0 |
| c9 | 0 | 0 | 0 | 0 | 0 | 0 | 0 | 0 | 0 | 0 | 0 | 0 | 0 | 0 | 0 |
| c10 | 0.5 | 0 | 0 | 0 | 0 | 0 | 0 | 0 | 0 | 0 | 0 | 0 | 0 | 0 | 0 |

o: firm (self), p: partner, g: local government, c: resident

Figure 2 illustrates the network graph constructed from the P/L adjacency matrix shown in Table 8. As in Figure 1, edges with saturated weights are drawn as thick lines, while edges with unsaturated weights are drawn as thin lines. The nodes $o1$ and $o2$ both represent the firm, that is, the same organizational entity.

Because the graph reflects only the current-period data, unlike in Figure 1, no social relationships are formed with residents $c4$, $c5$, and $c9$.

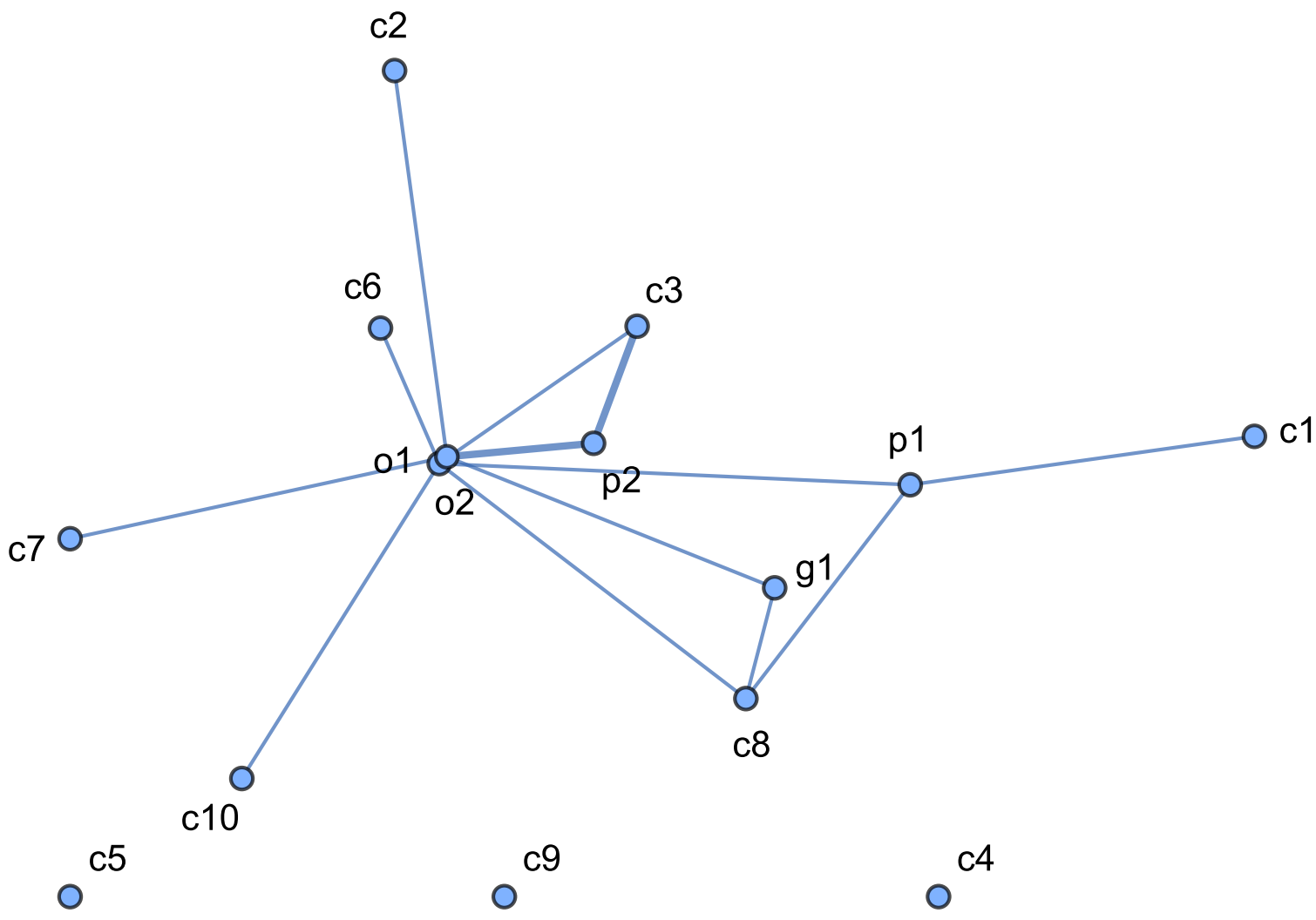

Figure 2. Network Graph Based on the P/L Adjacency Matrix.

Table 9 presents the profit–loss statement calculated from the P/L adjacency matrix, together with the managerial indicators derived using the formulas in Section 2.2. As with the balance sheet, conventional financial benchmarks cannot be directly applied; however, in this numerical example, all managerial indicators exhibit relatively high values.

Table 9. Profit–Loss Statement and Managerial Indicators Calculated from the P/L Adjacency Matrix.

<table>
<tr><th colspan="3">Profit–Loss Statement</th></tr>
<tr><td rowspan="5">Revenue<br>8.0</td><td colspan="2">Cost of Goods Sold<br>2.0</td></tr>
<tr><td colspan="2">Corporate Taxes<br>0.5</td></tr>
<tr><td>SG&A Expenses<br>2.5</td><td rowspan="3">Current Profit<br>5.5</td></tr>
<tr><td>Non-Operating Expenses<br>0.5</td></tr>
<tr><td>Net Income for the Period<br>2.5</td></tr>
</table>

| Managerial Indicators |
| --- |
| Net Profit Margin<br>0.69 |
| Return on Assets (ROA)<br>0.35 |
| Return on Equity (ROE)<br>0.52 |
| Return on Investment (ROI)<br>1.83 |
| — |

In this section, we demonstrated that social statements (the balance sheet and income statement) and managerial indicators can be computed by collecting data on social relationships, generating adjacency matrices, and visualizing the corresponding network graphs. Although simplified numerical examples were used for illustrative purposes, the computational procedures presented in this section remain applicable even when the numbers of partner, local-government, and resident nodes, as well as the number of edges among them, increase or decrease. This framework therefore enables the tracking of temporal changes in social relationships and their associated social value.

## 4. Discussion

In this study, we presented a set of social statements, formulated in parallel with conventional financial statements, based on the premise that social value is grounded in social relationships—namely, joint actions and communication. The proposed social statements offer several advantages: (1) they can be easily computed by collecting data on social relationships between the firm and external actors; (2) they allow temporal changes to be tracked through the accumulation of historical data; (3) the measurement units of the balance-sheet and income-statement items are unified, making the correspondence between the two statements explicit; and (4) as with financial statements, managerial indicators can be derived from the social statements and used for evaluative purposes.

In implementing the proposed social statements in real-world corporate settings, it is necessary to consider operational methods appropriate to different types of business activities. The main issues for further examination are outlined below.

- Measuring Social Relationships According to Business Type
  - For B-to-C firms such as those described in the Cooperative Business Survey Report [11] cited in the Introduction, producers and consumers constitute the primary stakeholders, making it relatively straightforward to measure social relationships between the firm and external individuals.
  - In B-to-B firms, direct contact with external individuals is limited. Thus, social relationships may need to be measured either as relationships between corporate entities or by considering downstream B-to-B-to-C interactions. In this case, social relationships with residents are mediated through partner firms, and the corresponding items appear primarily as liabilities in the balance sheet and as costs in the income statement.
  - Among B-to-C firms, differences arise between small retail shopping streets and large shopping centers. In the former, face-to-face customer interactions can be measured directly. In the latter, simply counting the number of people passing through checkout counters does not capture joint actions or bidirectional communication and therefore is not suitable for social statements. Additional measures—such as capturing interactions at the level of individual stores—may be required.
  - In essential services such as welfare and caregiving, the number of service recipients does not necessarily increase over time, and revenue (i.e., period-to-period increments) may remain small. However, if the support network surrounding service recipients is conceptualized as a social relationship, the social statements can still be meaningfully applied. For example, see the network analysis in Reference [15].
- Measuring Real and Virtual Social Relationships
  - Face-to-face social relationships in corporate activities can be measured using technologies such as RFID tags or proximity sensing via smartphones.
  - For virtual communication—email, chat, or social networking services—bidirectional exchanges can be treated as social relationships. In one-to-many advertising or broadcasting, a social relationship may be recognized when some form of response is generated.
- Correlation Between Social Impact and Social Statements
  - Although the social relationships used in the social statements do not represent social impact itself, larger values in the social statements indicate a greater number of sympathizers or collaborators in corporate activities—that is, greater recognition of social value.
  - Therefore, analyzing the correlation between social impact and the social statements or managerial indicators can provide multidimensional insights and support corporate decision-making.

Because the methods of measuring and operationalizing social statements vary depending on the nature of a firm's activities, comparing firms with different business models is of limited significance. The value of social statements lies instead in enabling firms to incorporate social value—rather than profit maximization alone—into their managerial decision-making, and in allowing firms to communicate their social value broadly to external stakeholders, thereby contributing to collective efforts toward addressing social problems.

As future work, it remains necessary to validate the proposed social statements in actual corporate settings and to examine their usefulness in managerial decision-making. Moreover, although not discussed in detail in this paper, various related initiatives have been developed, including natural-capital balance sheets and income statements [16], as well as the Social Return on Investment (SROI) framework, which converts social value into monetary terms [17]. In conjunction with these approaches, we hope that efforts to disseminate the concept and practice of social statements will contribute to broader social transformation.

## Acknowledgments

We received valuable insights regarding the challenges of corporate value and management metrics during the early stages of this research from Professor Yasuo Deguchi of the Graduate School of Letters at Kyoto University (currently the Institute for the Future of Humanities and Societies). We engaged in discussions regarding social and environmental statements—designed to correspond with financial statements—with Project Manager Koichi Watanabe and Senior Researchers Tsukasa Funane, Yosuke Tanabe, and Akinori Nishide of the Research and Development Group at Hitachi, Ltd. We also discussed social impact and social statements with Senior Researchers Ryuji Mine and Junichi Miyakoshi, as well as Researcher Misa Owa, from the same group. I would like to express my deepest gratitude to all of them.